\documentclass[fleqn,10pt]{wlscirep}
\usepackage[utf8]{inputenc}
\usepackage[T1]{fontenc}

\usepackage{newtxtext,newtxmath}
\usepackage{amssymb,mathtools}  % mathtools load amsmath
\usepackage{physics2}
\usephysicsmodule{ab, ab.legacy, op.legacy}
\usepackage{fixdif, derivative}
\usepackage{xcolor}
\usepackage{booktabs, tabularx, nicematrix}

\newcommand{\rmsub}[2]{#1_{\mathrm{#2}}}
\newcommand{\aet}{\aab{\tau_i}}
\newcommand{\met}{\overline{\tau}}
\newcommand{\nummethod}{Numerical methods and parameters}
\newcommand{\limitingcases}{Escape time in the two limiting cases}
\newcommand{\mfyisystem}{Changing variables to the mean field and displacements}
\newcommand{\doublefpe}{Approximate theory for weak coupling cases}

\title{Diffusive coupling facilitates and impedes noise-induced escape in interacting bistable elements}

\author[1,*]{Hidemasa Ishii}
\author[1]{Hiroshi Kori}
\affil[1]{Department of Complexity Science and Engineering, Graduate School of Frontier Sciences, the University of Tokyo, Chiba 277-8561, Japan}

\affil[*]{hidemasaishii1997@g.ecc.u-tokyo.ac.jp}

\begin{abstract}
  Diverse complex systems often undergo sudden changes in their states, such as epileptic seizures, climate changes, and social uprisings.
  Such behavior has been modeled by noise-induced escape of bistable elements, which is the escape from an attracting state driven by a fluctuation in the system's state.
  We consider a system of interacting bistable elements and investigate the effect of diffusive coupling among elements on the process of noise-induced escape.
  We focus on the influence of the coupling strength over the escape time, which is the time it takes for noise-induced escape to occur.
  We performed numerical simulations and observed that weak coupling reduced the mean escape time, whereas strong coupling impeded escape.
  We argue that, although diffusive coupling both facilitates and impedes escape, the facilitating effect is dominant when coupling is weak.
  For weak coupling cases, we develop an approximate theory that can predict the mean and variance of escape times.
  In contrast, strong coupling reduces the effective noise intensity to impede escape.
  Our results suggest that diffusive coupling among multistable elements contributes to regulating the rate of transitions among attracting states.
\end{abstract}
\begin{document}

\flushbottom
\maketitle
% * <john.hammersley@gmail.com> 2015-02-09T12:07:31.197Z:
%
%  Click the title above to edit the author information and abstract
%
\thispagestyle{empty}

\section*{Introduction}
\begin{figure}[tb]
  \centering
  \includegraphics[width=0.96\linewidth]{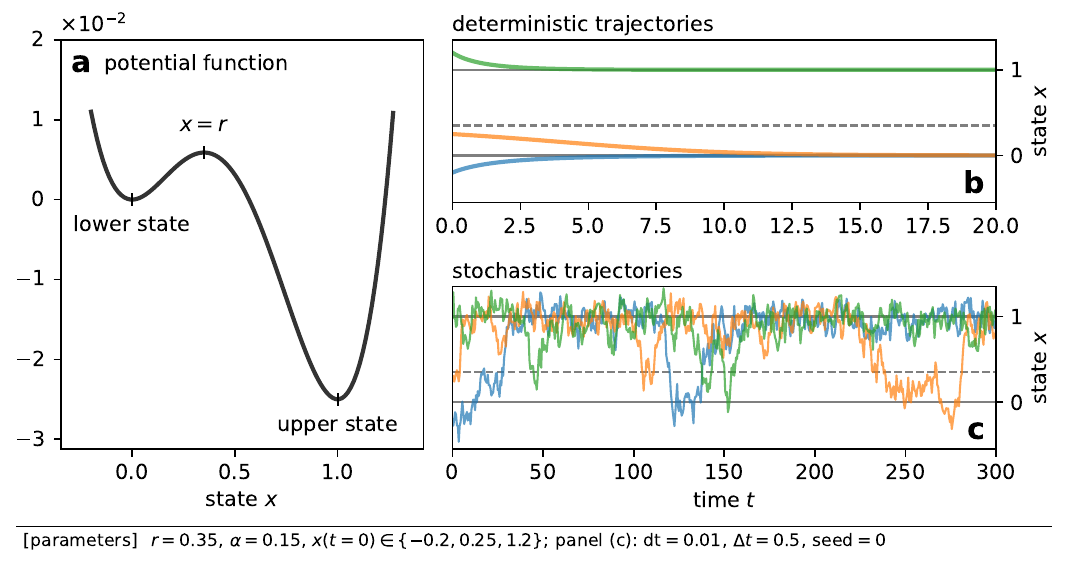}
  \caption{%
    While a state of a deterministic bistable element converges to an attracting state, noise-induced escape from an attracting state occurs when the system is subjected to noise.
    (a) Illustration of a bistable potential [equation~\eqref{eq:pot}] for $r < 1 / 2$.
    (b) Trajectories of an isolated deterministic bistable element, $\d{x} / \d{t} = f(x)$, with three initial conditions.
    (c) Trajectories of an isolated stochastic bistable element [equation~\eqref{eq:sde-uncoupled}] with three initial conditions.
  }
  \label{fig:demo}
\end{figure}

Epileptic brains~\cite{kalitzin2010Stimulationbased}, ecosystems~\cite{may1977Thresholds}, and firms adopting innovations~\cite{herbig1991cusp} --- such diverse complex systems exhibit the similar behavior, wherein their states undergo abrupt changes among multiple stable states.
This phenomenon has attracted much attention under the terms such as ``tipping points'', ``thresholds and breakpoints'', and ``regime shifts''~\cite{may2008Ecology}.
Bistable models, where a system has two distinct attracting states, play an essential role in studying such sudden transitions among different states.
Since complex systems often consist of a number of components that interact with each other, models of interacting bistable elements have been studied in diverse contexts, including epilepsy~\cite{benjamin2012phenomenological,lopes2023role}, abrupt change in ecosystems~\cite{dakos2010Spatial}, climate change~\cite{wunderling2021Interacting}, poverty traps~\cite{barrett2006Fractal}, and the spread of uprising during the Arab Spring~\cite{brummitt2015Coupled}.
A pile of theoretical literature also exist, for instance on the influence of underlying network structures~\cite{kronke2020Dynamics} and the prediction of tipping points~\cite{scheffer2009Earlywarning,kundu2022Meanfield,maclaren2023Early}.

When a bistable system is deterministic, its two attracting states are also stationary states.
Hence, the system converges into one of the two states [figure~\ref{fig:demo}(b)].
In deterministic cases, the interaction with other bistable elements causes the propagation of one stationary state and also the coexistence of the two stationary states among elements~\cite{booth1992Mechanisms,erneux1993Propagating,kouvaris2012Traveling,caputo2015Bistable}.
The propagation and coexistence have been observed empirically in mechanical systems~\cite{nadkarni2016Unidirectional,raney2016Stable} and electrochemical reactions~\cite{kouvaris2016SelfOrganized,kouvaris2017Stationary}.
When a bistable system is stochastic, its state not only fluctuates about an attracting state but also transitions between the two attracting states intermittently [figure~\ref{fig:demo}(c)].
Because the escape from an attracting state is driven by noise, it is called noise-induced escape.
In stochastic cases, the interaction among bistable elements affects the rate of noise-induced escape~\cite{frankowicz1982Stochastic,neiman1994Synchronizationlike,benjamin2012phenomenological,ashwin2017Fast,creaser2018Sequential}.

A transition among attracting states corresponds to a dramatic change in the system's state, such as an epileptic seizure, extinction of a species, and a riot.
Therefore the mean escape time, which is the time it takes for noise-induced escape to occur, is of particular interest.
Among the literature on systems of coupled stochastic bistable elements, Frankowicz and Gudowska-Nowak~\cite{frankowicz1982Stochastic} reported in 1982 that, while weak diffusive coupling reduced the mean escape time, stronger coupling slowed down escape.
Recently, Creaser and the colleagues~\cite{creaser2018Sequential} also briefly referred to this non-monotonic dependence of the mean escape time on the coupling strength.
Previous research has mainly studied small systems with $2$ or $3$ elements, as they are analytically tractable and computationally less demanding.
Considering that many real systems --- such as brains, ecosystems, and society --- consist of a number of components, the effect of the coupling strength on the mean escape time need be investigated for larger systems as well.
For small systems, several studies~\cite{ashwin2017Fast,creaser2018Sequential} succeeded in estimating the mean escape time utilizing the multidimensional Kramers' formula.
However, a similar analytical approach deems infeasible for larger systems due to their large degrees of freedom.
Moreover, even though the formula mathematically explains the relation between the coupling strength and the mean escape time, we still lack an intuitive understanding on the influence of the interaction among bistable elements over the process of noise-induced escape.

In this research, we consider a larger system of interacting bistable elements.
We assume diffusive coupling among elements, i.e. each element is affected by the difference between it's state and others' states.
Diffusive coupling has been assumed in the literature on noise-induced escape in systems of interacting bistable elements~\cite{frankowicz1982Stochastic,ashwin2017Fast,creaser2018Sequential}, and on pattern formation in networked bistable reaction-diffusion models~\cite{booth1992Mechanisms,erneux1993Propagating,kouvaris2012Traveling,caputo2015Bistable,kouvaris2016SelfOrganized,kouvaris2017Stationary}.
In addition, it has been employed to model mechanical systems~\cite{nadkarni2016Unidirectional,raney2016Stable}, epileptic brain~\cite{benjamin2012phenomenological}, and ecosystem~\cite{dakos2010Spatial}.

Our first result is the relation between the coupling strength and the numerically measured mean escape time.
Direct numerical simulations revealed that weak coupling accelerates escape on average, while strong coupling impedes escape.
We then discuss the role of diffusive coupling in the process of noise-induced escape.
As mentioned above, an analytical approach similar to the previous studies is infeasible.
Instead, we describe how weak and strong coupling would change the behavior of each bistable element.
Whereas diffusive coupling to a node that has already escaped facilitates escape, interaction with a node that has not escaped impedes escape.
When coupling is weak, the balance between these facilitating and impeding effects determines the mean escape time.
For weak coupling cases, we develop an approximate theory that predicts the mean and variance of escape times.
As coupling becomes stronger, the effective noise intensity for the system declines.
We discuss the scaling of the effective noise intensity for strong coupling cases.
To facilitate analyses, our model assumes the global coupling among elements and the asymmetric bistability, where the stabilities of the two attracting states differ.
Still, we expect that coupling affects the process of noise-induced escape in a qualitatively similar manner on other network structures or with the symmetric bistability.
This research thus offers fundamental insights into the role of diffusive coupling in systems of interacting multistable elements, which have been employed in diverse fields from biology to social sciences.

\section*{Model}
\subsection*{Dynamics of each element}
First, we introduce the basic model that describes the dynamics of each element.
When we assume no interaction among elements, our model reduces to the following stochastic differential equation (SDE),
\begin{equation} \label{eq:sde-uncoupled}
  \d x = f(x) \d t + \alpha {\d W(t)},
\end{equation}
where
\begin{equation} \label{eq:schloegl}
  f(x) \coloneq -x (x - r) (x - 1),
\end{equation}
$x$ denotes the state of one element, $\alpha$ is the noise intensity, and $W(t)$ is a standard Wiener process.
$f(x)$, which was introduced by Schl{\"o}gl~\cite{schlogl1972Chemical}, describes the deterministic bistable dynamics of each element.
Its double-well potential
\begin{equation} \label{eq:pot}
  V(x) \coloneq \frac{1}{4} x^4 - \frac{1 + r}{3} x^3 + \frac{r}{2} x^2,
\end{equation}
which satisfies $f(x) = -\d V(x) / \d x$, is plotted in figure~\ref{fig:demo}(a).
The deterministic dynamics $\dot{x} = f(x)$ has two attracting states located at the two minima of $V(x)$.
The lower attracting state is at $x = 0$, and the upper one is at $x = 1$.
Parameter $r \in (0, 1)$ controls the asymmetry of the potential.
That is, the relative stability of the upper state against the lower one increases as $r$ gets smaller.
We assume $r$ is small, in which case the transition from the upper to lower state is rare enough to be negligible.

\subsection*{Globally coupled stochastic bistable elements}
In this paper, we consider a globally coupled network whose each node is the stochastic bistable element.
The dynamics of the system is governed by the following SDE,
\begin{equation} \label{eq:sde-full}
  \d x_i = \ab[f(x_i) + \frac{K}{N} \sum_{j = 1}^{N} \ab(x_j - x_i)] \d t + \alpha \d W_i(t),
\end{equation}
where $x_i$ denotes the state of node $i$, $K$ is the coupling strength, $N$ is the number of nodes in a system, and $W_i(t)$ are independent Wiener processes.
Since we assume the global coupling, one can rewrite the model \eqref{eq:sde-full} as
\begin{equation} \label{eq:sde-w/mf}
  \d x_i = \ab[f(x_i) + K \ab(X - x_i)] \d t + \alpha \d W_i(t),
\end{equation}
where $X$ is the mean field,
\begin{equation} \label{eq:mfdef}
  X \coloneq \frac{1}{N} \sum_{j = 1}^{N} x_j.
\end{equation}

\subsection*{Mean escape time}
This research investigates the system's escape from $x = 0$ to $x = 1$, focusing on the mean escape time.
We initialize the system to the lower state ($x = 0$) and analyze the time it takes for the system to escape to the upper one ($x = 1$).
Technically, the first escape time of node $i$ is defined as
\begin{equation}
  \tau_i \coloneq \inf_{t} \ab\{t > 0 \text{ such that } x_i(t) \geq \xi \text{ given } x_i(0) = 0 \},
\end{equation}
where $\xi$ is a fixed threshold between the lower and upper states.
The value of $\xi$ is arbitrary as long as it is not close to $r$ or $1$.
We chose $\xi = 0.5$ in the following, but our results remain valid for other $\xi$ values.
As $\tau_i$ is defined for each node, we also define the average escape time,
\begin{equation}
  \aet \coloneq \frac{1}{N} \sum_{i = 1}^{N} \tau_i.
\end{equation}
The average $\aab{\cdot}$ is over nodes, not noise realizations.
In other words, $\aet$ is defined for each sample path.
By taking expectation, we obtain the expected average escape time
\begin{equation}
  \met \coloneq \mathbb{E} \ab[\aet],
\end{equation}
which we call mean escape time in this article.
When the system is one dimensional, one can employ the following formula for mean escape time $T$~[Section 5.5 in Ref.~\cite{gardiner2009Stochastic}]:
\begin{equation} \label{eq:metformula}
  T(\alpha) = \frac{2}{\alpha^2} \int_0^{\xi} \d y \exp \ab(\frac{V(y)}{\alpha^2 / 2}) \int_{-\infty}^{y} \d z \exp \ab(-\frac{V(z)}{\alpha^2 / 2}).
\end{equation}
In addition to equation~\eqref{eq:metformula}, Kramers' theory~\cite{kramers1940Brownian,berglund2013Kramers} is often employed to study noise-induced escape~\cite{neiman1994Synchronizationlike,ashwin2017Fast,creaser2018Sequential}.
The Kramers' formula for the mean escape time is the approximation of the formula~\eqref{eq:metformula} in the weak noise limit, and expressed as
\begin{equation} \label{eq:kramers}
  \tilde{T}(\alpha) = \frac{2 \pi}{\sqrt{V''(0) \abs{V''(r)}}} \exp \ab(\frac{V(r) - V(0)}{\alpha^2 / 2}).
\end{equation}

\section*{Results}
\subsection*{Numerically measured mean escape time}
\begin{figure}[tbp]
  \centering
  \includegraphics[width=.96\linewidth]{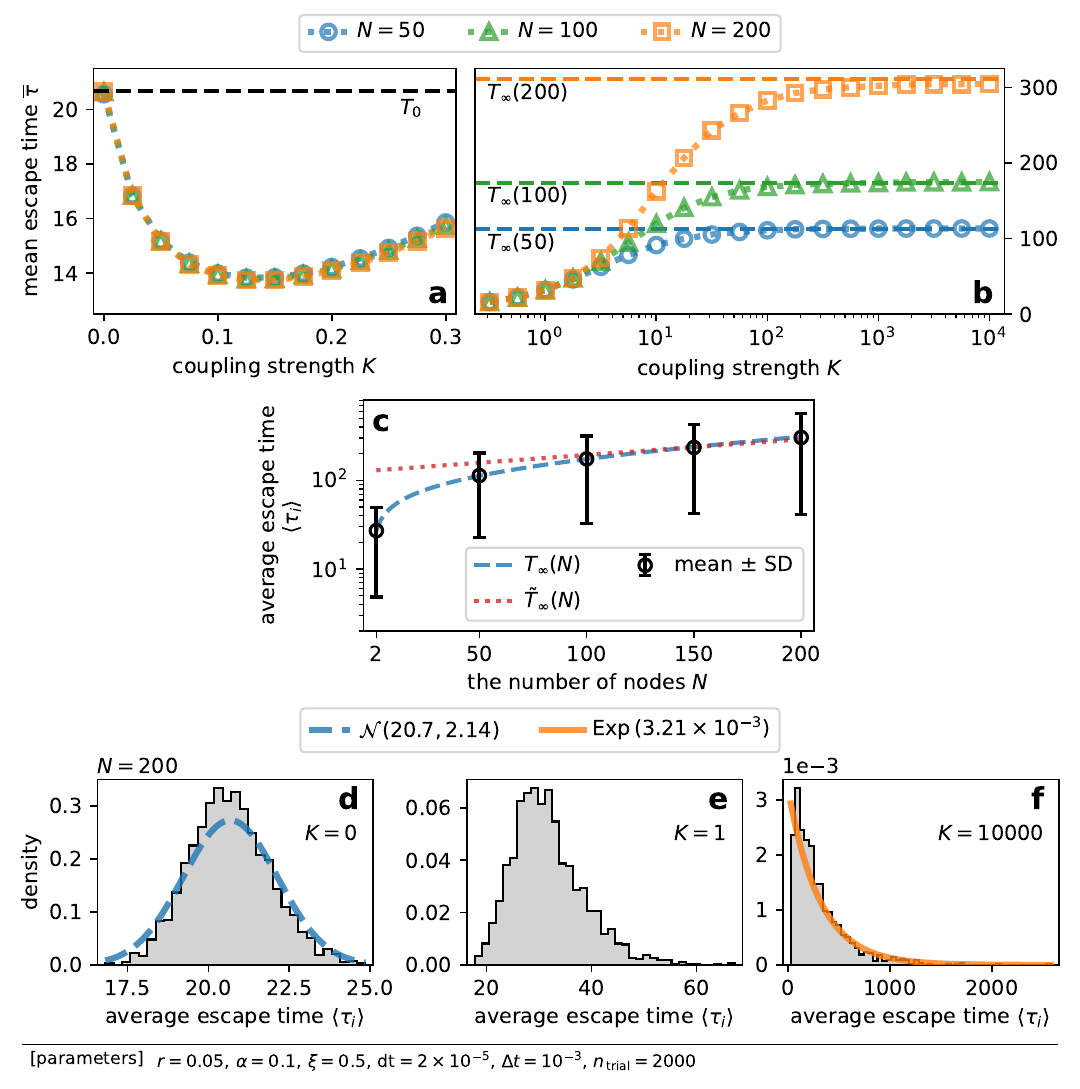}
  \caption{%
    The direct numerical simulations of the model [equation~\eqref{eq:sde-w/mf}] indicated that weak coupling reduced the mean escape time, while strong coupling impeded escape.
    (a, b) Numerically obtained mean escape time $\met$ against the coupling strength $K$ for $N = 50$, $100$, and $200$.
    (c) The system size dependence of average escape time $\aet$ in the strong coupling limit.
    Black markers show the mean escape time $\met$, and error bars indicate standard deviations (SD).
    The blue dashed line is the prediction of equation~\eqref{eq:tinfty}, whereas the red dotted line indicates Kramers' formula [equation~\eqref{eq:kramersinfty}].
    (d -- f) Histogram of average escape times for $K = 0$, $1$ and $10000$.
    The dashed and solid lines show the theoretical probability distribution functions of the normal [$\mathcal{N}(T_0, {T_0}^2 / N)$] and exponential [$\mathrm{Exp}(1 / T_{\infty}(N))$] distributions.
    }
  \label{fig:res-all}
\end{figure}

Figures~\ref{fig:res-all}(a, b) plot the mean escape time against the coupling strength for $N = 50$, $100$, and $200$.
The mean escape time was measured by direct numerical simulations of the model SDE~\eqref{eq:sde-w/mf}.
The figures exhibit the similar trends to the literature on a small system~\cite{frankowicz1982Stochastic}.
That is, while weak coupling accelerated noise-induced escape, strong coupling impeded escape.

There are two trivial limiting cases.
% K = 0
First, when there is no interaction, i.e. $K = 0$, all nodes are independent of each other.
In this case, one expects the mean escape time to be
\begin{equation} \label{eq:t0}
  T_0 \coloneq T(\alpha),
\end{equation}
which does not depend on the system size $N$.
Figure~\ref{fig:res-all}(a) shows the mean escape time at $K = 0$ indeed coincided with $T_0$ for all $N$.
% K to infty
Second, in the limit of strong coupling, i.e. $K \to \infty$, differences among nodes' states decay so fast that one may assume $x_i \approx X$.
The system reduces to the following one-dimensional system for the mean field $X$:
\begin{equation}
  \d{X} = f(X) \d{t} + \frac{\alpha}{\sqrt{N}} \d{W_X}(t),
\end{equation}
where $W_X(t)$ is a standard Wiener process.
As the system is one-dimensional, one can employ the formulae for the mean escape time~[equations~\eqref{eq:metformula} and \eqref{eq:kramers}] with the reduced noise intensity $\alpha / \sqrt{N}$:
\begin{gather} \label{eq:tinfty}
  T_{\infty}(N) \coloneq T\ab(\alpha / \sqrt{N}), 
  \\ \label{eq:kramersinfty}
  \tilde{T}_{\infty}(N) \coloneq \tilde{T}\ab(\alpha / \sqrt{N}).
\end{gather}
Indeed, figure~\ref{fig:res-all}(b) demonstrates that the mean escape time saturated to approach $T_{\infty}(N)$ as $K$ increased.
In addition, the prediction of equation~\eqref{eq:tinfty} was validated in figure~\ref{fig:res-all}(c).
Figure~\ref{fig:res-all}(c) also shows Kramers' formula became relevant as $N$ increased.
This is because Kramers' formula is valid in the weak noise limit.
We give more explanations on the two limiting cases in ``{\limitingcases}'' in Methods section.

\subsection*{Weak coupling facilitates escape}
\begin{figure}[tb]
  \centering
  \includegraphics[width=0.96\linewidth]{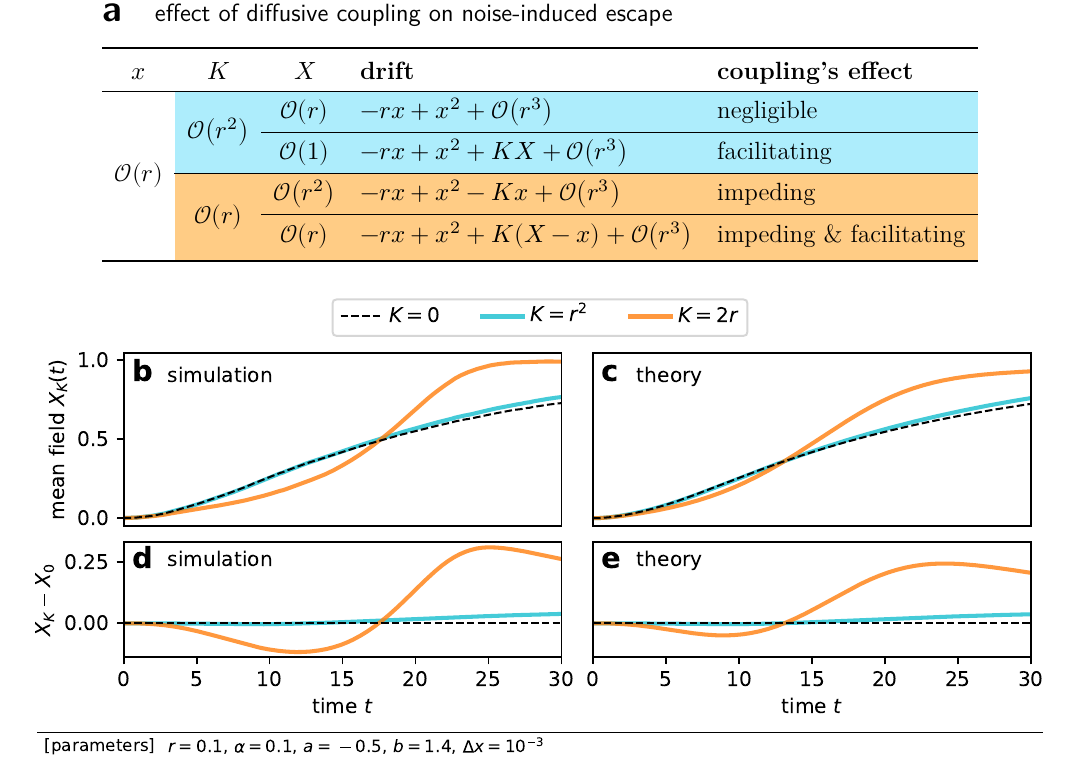}
  \caption{%
    The facilitating effect of diffusive coupling is dominant when coupling is sufficiently weak.
    (a) The leading terms of the drift term of the model SDE~\eqref{eq:sde-w/mf}.
    Coupling may facilitate and impede noise-induced escape depending on the order of coupling strength $K$ and the mean field value $X$.
    $x = \order{r}$ is assumed because we are interested in nodes that have not escaped.
    (b, c) Trajectories of $X_K(t)$, the mean field when the coupling strength is $K$.
    (d, e) The difference of $X_K(t)$ from $X_0(t)$ as a function of time.
    Panels (b) and (d) are the results of direct numerical simulations of the model SDE.
    Coupling was so weak that nodes were almost independent of each other.
    The trajectories of the mean field are thus nearly independent of noise realizations.
    Panels (c) and (e) are the results of our approximate theory.
  }
  \label{fig:weak-mfcomp}
\end{figure}

\begin{figure}[tb]
  \centering
  \includegraphics[width=0.96\linewidth]{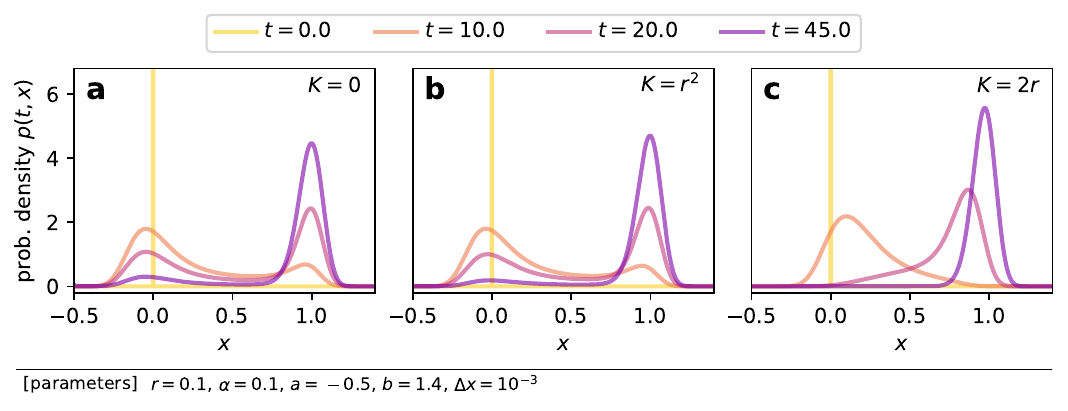}
  \caption{Evolution of the probability density function $p(t, x)$ according to our approximate Fokker-Planck equation~\eqref{eq:fpe-k}.}
  \label{fig:weak-pdf}
\end{figure}

\begin{figure}[tb]
  \centering
  \includegraphics[width=0.96\linewidth]{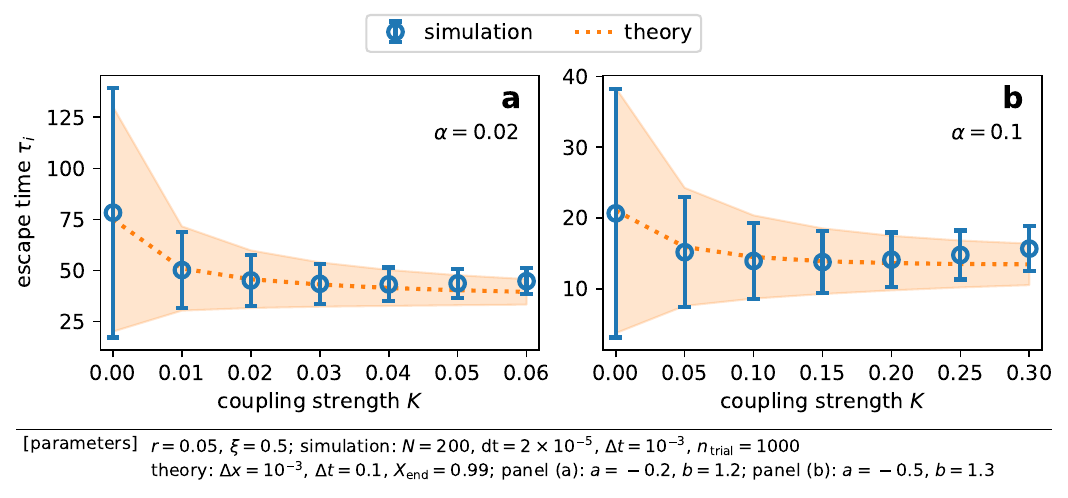}
  \caption{%
    Our theory succeeded in estimating the mean and variance of escape times for different parameter values.
    (a, b) Comparison of the mean and standard deviation (SD) of escape times between direct simulations and our theory.
    Markers and error bars show respectively the mean escape time and the SD of escape times among nodes obtained through direct simulations.
    The dotted line and the filled area respectively indicate the mean and SD predicted by our theory.
  }
  \label{fig:met-sd-comp}
\end{figure}

Figure~\ref{fig:res-all}(a) revealed weak coupling reduced the mean escape time with the minimum around $K \approx 0.15$.
Since the coupling in our model is diffusive, it brings about the synchronization of nodes' states.
That is, the stronger the coupling is, the more aligned each node's state is to the mean field.
Therefore the coupling impedes escape when the mean field is small ($X \approx 0$), and it facilitates escape when $X$ is large.
When the coupling is very weak, the latter facilitating effect is dominant, which is why weak coupling reduced the mean escape time.
In this subsection, we elaborate on the influence of weak coupling over noise-induced escape.

Figure~\ref{fig:weak-mfcomp}(a) presents the temporal evolution of the mean field for several $K$ values.
The trajectories were obtained by numerical integration of the model SDE~\eqref{eq:sde-w/mf}.
One sees the mean field $X$ monotonically increased, because we assumed $r$ was so small that we could neglect escape from the upper ($x = 1$) to lower ($x = 0$) state.
This implies that the impeding effect would be dominant in the early period when $X$ is small, and the facilitating effect would occur later.

To elucidate how weak coupling affects noise-induced escape, we focus on a node that has not yet escaped.
In particular, we assume the order of the node's state $x$ is $\order{r}$ [figure~\ref{fig:weak-mfcomp}(a)].
In this case, the leading term of $f(x)$ is $\order{r^2}$, i.e. $f(x) \sim r^2$.
When coupling is weak enough to satisfy $K \sim r^2$, the term $-Kx \sim r^3$ is negligible compared with the isolated dynamics $f(x)$.
Moreover, when the mean field $X$ is small enough to satisfy $X \sim r$, we have $KX \sim r^3$, implying that the whole coupling term $K(X - x)$ is negligible with regard to $f(x)$.
After some time, the mean field $X$ grows to be $\order{1}$.
Then, whereas the term $-Kx \sim r^3$ is still negligible, $KX \sim r^2$ becomes comparable to $f(x)$.
The coupling term reduces as $K(X - x) \simeq KX > 0$, indicating that the coupling's dominant effect on escape is facilitating.
In the case of $K \sim r$, i.e. less weak coupling, the $-Kx$ term is no longer negligible because $-Kx \sim r^2 \sim f(x)$.
Furthermore, in the initial period where $X \sim r^2$, one may ignore $KX \sim r^3$ to obtain the effective coupling term of $K(X - x) \simeq -Kx$, which would be negative on average and thus impede escape.

Our argument is summarized in figure~\ref{fig:weak-mfcomp}(a).
When coupling is sufficiently weak to satisfy $K \sim r^2$, we expect ({\romannumeral 1}) the dynamics are almost the same as that in the no coupling case ($K = 0$) in the early period, and ({\romannumeral 2}) only the facilitating effect occurs after the growth of the mean field.
When the coupling strength is increased to become $K \sim r$, we also expect ({\romannumeral 3}) coupling impedes escape in the initial period.
These expectation were confirmed in figures~\ref{fig:weak-mfcomp}(b, d).
First, the trajectories for $K = 0$ and $r^2$ overlap until around $t \approx 15$.
Indeed, the difference between $X_K$ for $K = r^2$ and $X_0$ [figure~\ref{fig:weak-mfcomp}(d)] remained close to $0$ in the early period.
Second, the difference $X_K - X_0$ increased after the early period.
The increase in the difference corresponds to the faster growth of $X_K$ than $X_0$, illustrating coupling's facilitating effect.
Third, comparing the cases of $K = 0$ and $2r$, one finds that the growth of $X_K$ was slower than $X_0$ in the early period, which demonstrates coupling's impeding effect.

To obtain quantitative insights into the mean escape time in weak coupling cases, we developed an approximate theory.
Our arguments above imply that the initial evolution of the mean field is close to that in the uncoupled case when coupling is weak.
We thus assume $X(t) \simeq X_0(t)$, where $X_0(t)$ is the trajectory of the mean field when $K = 0$.
Substituting $X_0(t)$ for $X$ in the model SDE~\eqref{eq:sde-w/mf}, our model becomes a one-dimensional system for $x$ with the time-dependent parameter $X_0(t)$, for which one can solve the Fokker-Planck equation (FPE).
The trajectory of $X_0(t)$ is also available by solving the FPE for the uncoupled model~\eqref{eq:sde-uncoupled}.
Hence, by simultaneously solving the FPEs for uncoupled and weakly coupled models, we obtain the probability density function $p(t, x)$ for $x$ at time $t$ in weak coupling cases.
One can furthermore compute the probability density function for escape times from $p(t, x)$.
We refer readers to ``\doublefpe'' in Methods section for details.

Figures~\ref{fig:weak-mfcomp}(c, e) present the estimated trajectories of the mean field $X_K(t)$ computed by our approximate theory.
Comparing with the results from the direct numerical simulations [figures~\ref{fig:weak-mfcomp}(b, d)], trajectories for $K = 0$ and $r^2$ seem almost identical, and the theoretical curves in figures~\ref{fig:weak-mfcomp}(c, e) deviated from those in figures~\ref{fig:weak-mfcomp}(b, d) for $K = 2r$.
The difference between the actual $X(t)$ and $X_0(t)$, which is neglected in our theory, is naturally the cause of the deviation for $K = 2r$.
Figure~\ref{fig:weak-pdf} shows the snapshots of the probability density functions.
The upper peak around $x = 1$ at $t = 45$ was higher for stronger coupling, indicating the acceleration of the collective escape process.
At $t = 10$ and $20$, one finds two peaks around $x = 0$ and $1$ for $K = 0$ [figure~\ref{fig:weak-pdf}(a)] and $r^2$ [figure~\ref{fig:weak-pdf}(b)], but the distribution was no longer bimodal for $K = 2r$ [figure~\ref{fig:weak-pdf}(c)].
This demonstrates the synchronizing effect of diffusive coupling.
Figure~\ref{fig:met-sd-comp} presents our main results for weak coupling cases, where the estimate from our theory is compared with the results of direct numerical simulations.
By solving FPEs, one obtains not only the mean but also the variance of escape times $\tau_i$, which are depicted by the dotted line and the area plot.
Our estimates agreed surprisingly well to the simulation results as long as coupling was weak.

\subsection*{Strong coupling reduces the effective noise intensity}
\begin{figure}[tb]
  \centering
  \includegraphics[width=0.96\linewidth]{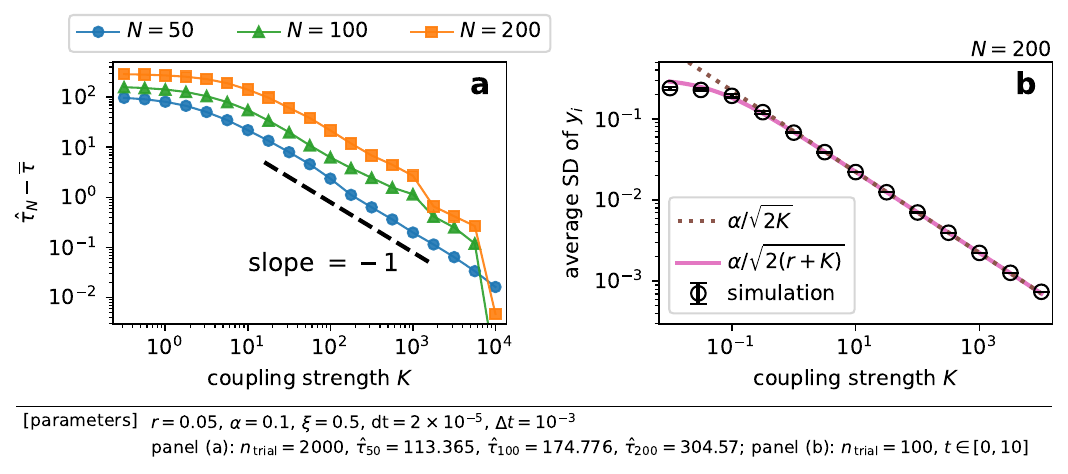}
  \caption{%
    Strong coupling reduces the effective noise intensity, resulting in slow escape.
    (a) The difference of the numerically measured mean escape time $\overline{\tau}$ from the asymptotic value $\hat{\tau}_N$.
    (b) The average standard deviation (SD) of displacements $y_i$ during $t \in [0, 10]$.
    The variance $\mathrm{Var}[y_i] \simeq \alpha^2 / [2 (K - f'(X))]$ is approximately $\alpha^2 / [2 (K + r)]$ when $X \ll 1$, which is shown by the solid line.
    When $r \ll K$, we obtain equation~\eqref{eq:ou-var}, which is indicated by the dotted line.
    The markers and error bars represent the mean and standard deviation of the simulation results.
  }
  \label{fig:strong}
\end{figure}

One sees from figures~\ref{fig:res-all}(a, b) that the mean escape time approached to the asymptotic value which was close to $T_{\infty}(N)$.
We manually determined the asymptotic value, which is denoted by $\hat{\tau}_N$, and computed the difference of the mean escape time from $\hat{\tau}_N$.
The result is presented in figure~\ref{fig:strong}(a), where the difference $\hat{\tau}_N - \met$ is plotted against the coupling strength $K$.
The figure illustrates that the difference scaled as $K^{-1}$ when $K$ was large.
We argue that this is because the effective noise strength is reduced as $K^{-1}$ as the coupling gets stronger.

To analyze the case of the finite coupling strength, i.e. $K \in (0, \infty)$, we introduce 
\begin{equation}
  y_i \coloneq x_i - X,
\end{equation}
which is the displacement of node $i$ from the mean field $X$.
Neglecting higher order terms $\order{y_i^2}$, the following equations describe the dynamics of $X$ and $y_i$,
\begin{align}
  \label{eq:mfyireduced-x}
  \d{X} =& f(X) \d{t} + \frac{\alpha}{\sqrt{N}} \d{W_X}(t), \\
  \label{eq:mfyisreduced-yi}
  \d{y_i} =& -\ab[K - f'(X)] y_i \d{t} + \alpha \sqrt{\frac{N - 1}{N}} \d{\tilde{W}_i}(t), && i \in \ab\{1, \dots, N - 1\},
\end{align}
where $W_X$ and $\tilde{W}_i$ are independent standard Wiener processes.
Their derivations are in ``{\mfyisystem}'' in Methods section.
Unless $K$ is so small that $f'(X) \geq K$, equation~\eqref{eq:mfyisreduced-yi} describes the Ornstein-Uhlenbeck process whose mean is zero.
When $N$ is large and $K \gg f'(X)$, the variance of $y_i$ would be approximately 
\begin{gather} \label{eq:ou-var}
  \mathrm{Var}[y_i] \simeq \frac{\alpha^2}{2K}
\end{gather}
after a short [$\order{K^{-1}}$] transient period.
Equation~\eqref{eq:ou-var} implies that, the stronger the coupling is, the closer the nodes are to the mean field $X$.
In other words, strong coupling enhances synchronization among nodes, reducing the effective degree of freedom of the system.
The results of numerical simulations shown in figure~\ref{fig:strong}(b) illustrate the decline in the standard deviation of $y_i$ as $K$ increased, verifying equation~\eqref{eq:ou-var}.

Remembering $x_i = X + y_i$, equations~\eqref{eq:mfyireduced-x} and \eqref{eq:mfyisreduced-yi} imply that $x_i$ is always subjected to noise whose intensity is $\alpha / \sqrt{N}$ regardless of the coupling strength.
This originates from the mean field dynamics [equation~\eqref{eq:mfyireduced-x}].
The second term $y_i$ is the Ornstein-Uhlenbeck process whose mean is $0$ and variance is about $\alpha^2 / (2K)$.
Therefore we expect the effective noise intensity for $x_i$ to be approximately
\begin{equation} \label{eq:effective-noise}
  \sqrt{\frac{\alpha^2}{N} + c' \frac{\alpha^2}{2K}} \simeq \frac{\alpha}{\sqrt{N}} \ab(1 + \frac{c}{K})
\end{equation}
for large $N$ and $K$, where $c$ is some constant.
This is why the difference $\hat{\tau}_N - \met$ scaled as $K^{-1}$.
Indeed, if one fixes the value of $c$ and substitutes equation~\eqref{eq:effective-noise} into equation~\eqref{eq:metformula}, one obtains a similar curve to the ones shown in figure~\ref{fig:strong}(a).
We note however that we have not been able to systematically determine the coefficient $c$.
This task is challenging, mainly because one must perform some kind of white approximation of the Ornstein-Uhlenbeck process $y_i$ in order to derive the form $\d{x} = \cdots + \alpha / \sqrt{N} (1 + c / K) \d{\rmsub{W}{effective}}(t)$.

\section*{Discussion}
We studied the effect of the coupling strength on noise-induced escape for a system of globally coupled bistable elements.
We numerically measured the mean escape time to observe that weak coupling reduced the mean escape time, whereas stronger coupling impeded escape [figures~\ref{fig:res-all}(a, b)].
We explained how weak coupling accelerates escape on average.
Although diffusive coupling both facilitates and impedes escape, only the facilitating effect is dominant when coupling is weak, resulting in the decline in the mean escape time (figure~\ref{fig:weak-mfcomp}).
Based on this idea, we succeeded in estimating the mean and variance of escape times in weak coupling cases (figure~\ref{fig:met-sd-comp}).
Finally, we reported that the difference of the mean escape time from its asymptotic value at $K \to \infty$ scaled as $K^{-1}$ [figure~\ref{fig:strong}(a)], which is due to the reduction in the effective noise intensity.

The phenomenon that weak coupling accelerates and strong coupling impedes noise-induced escape was reported for a two-node system in as early as 1982~\cite{frankowicz1982Stochastic}.
Our results revealed their observations could be extended to larger systems.
We furthermore gave an intuitive explanation for the way in which diffusive coupling affects the process of noise-induced escape.
Coupling enables escaped nodes to pull others out from the lower state to facilitate escape, but at the same time allows nodes that have not escaped to gather around the lower state to impede escape.
In addition, stronger coupling reduces the effective noise intensity.
When coupling is sufficiently weak, the first facilitating effect is dominant.
As coupling gets stronger, the second impeding effect becomes no longer negligible, and the balance between the facilitating and impeding effects determines the mean escape time.
Under stronger coupling, all nodes fluctuate around the mean field, and the reduction in the effective noise intensity leads to the slow escape.

% N = \order{100} is large enough because we have no reason to expect something different would occur for larger systems.

% on diffusive coupling
This research assumed that interaction among elements was diffusive.
Another popular choice is additive coupling, which assumes each element is affected by the sum of its neighbors' states.
Since additive and diffusive coupling can lead to different dynamics, the appropriate form of coupling must be determined when constructing a model~\cite{lopes2023role}.
Nonetheless diffusive coupling is of relevance to diverse topics.
Diffusive coupling is a reasonable assumption when interaction involves flow of substances or individuals such as plants and animals.
Among neurons, gap junction~\cite{trenholm2019Myriad} corresponds to diffusive coupling.
We also believe diffusive coupling can describe social interaction among persons.
Diverse states and behavior of human individuals, such as smoking~\cite{ali2009Estimating} and emotions~\cite{kramer2014Experimental}, are known to spread through influence of peers, which is termed social contagion~\cite{christakis2013Social}.
Social learning through imitation~\cite{whiten2009Emulation,chartrand2013Antecedents,tamura2015Evolution} is a salient mechanism by which behavior of an individual is affected by others.
When an individual imitates those who are less similar, the magnitude of change in the person's state would be greater.
This illustrates resemblance between social interaction through imitation and diffusive coupling.
We note, even if additive coupling is assumed in our model, weak coupling would still reduce the mean escape time.
The additive coupling term under global coupling is $K / N \sum_j x_j = K X$, which is generally positive and thus facilitates escape.

% symmetric case
In addition to diffusive coupling, this study assumed the asymmetric bistability, where the upper state is much more stable than the lower one.
Nevertheless the effect of the coupling strength on the mean escape time would be qualitatively similar to our results when the potential is symmetric.
That is, weak coupling would facilitate noise-induced escape, and strong coupling would reduce the effective noise intensity to impede escape.
The symmetric case is especially relevant in studying stochastic resonance~\cite{gammaitoni1998Stochastic}.
When the coupling strength is fixed to a large value, one would observe stochastic resonance by changing the system size, as the effective noise intensity is about $\alpha / \sqrt{N}$.
Indeed, such phenomena is known as system size resonance~\cite{pikovsky2002System}.
Moreover, since the transition rate changes according to the coupling strength, one also observes stochastic resonance by changing the coupling strength, instead of the noise intensity.
This coupling-induced stochastic resonance may allow for the empirical estimation of the coupling strength.
For instance, one may conduct stochastic resonance experiments for a system of interacting elements~\cite{gluckman1996Stochastic} and its isolated element.
Because weak (strong, respectively) coupling increases (reduces) the transition rate, one expects the interaction within the system to be weak (strong) if the optimal noise intensity for the whole system is smaller (larger) than that for an isolated element.

Our results suggest that the diffusive coupling among multistable elements contributes to regulating the transition rate among attracting states.
When each element is subjected to noise, interaction with others changes the rate of noise-induced escape.
Strong coupling would be appropriate for a system where the high stability of a particular attracting state is favorable, since it reduces the transition rate.
In contrast, weak coupling can improve the efficiency of some process by facilitating escape.
It is particularly interesting if weak coupling promotes stochastic resonance that is unattainable without the coupling.
Putting it another way, when the environmental noise is too weak to induce stochastic resonance, weak coupling may amplify the effective noise to achieve the optimum noise intensity for resonance.
Previous studies have argued that stochastic resonance enhances sensory capacities~\cite{russell1999Use,itzcovich2017Stochastic}.
Our results indicate the possibility that relevant systems, perhaps neuronal systems, evolved to have weak coupling among their components so that they can exploit stochastic resonance.

% limitations / open questions
We restricted ourselves to studying the expected average escape time, which we referred to as the mean escape time.
Depending on the context, other quantities may well be used to characterize the escape process of the whole system.
Our results would be qualitatively valid even if one adopts different quantities such as the expected escape time of the last node $\mathbb{E}[\tau_N]$ and the expected median of escape times.
However, it is no longer the case if one focuses on the expected escape time of nodes that escape early, for instance $\mathbb{E}[\tau_1]$.
Indeed, the expected escape time of the first node $\mathbb{E}[\tau_1]$, also known as extreme first passage time~\cite{lawley2020Distribution}, would monotonically increase as the diffusive coupling becomes stronger, because the coupling reduces the variance in escape times among nodes.

% - on general networks
While this research is limited to the global coupling case, the influence of the coupling strength over the mean escape time would be qualitatively similar for other network structures.
For general network cases, it might be possible to perform similar analyses to ours by introducing the mean field weighted by nodes' degrees, instead of the simple mean field $X$.
% - heterogeneity in potential
Another interesting direction for future research is the introduction of heterogeneity.
In various real systems, the shape of the potential determined by $r$ is likely to differ across elements.
The coupling strength may also vary across interactions, i.e. across edges in a network.
Our work lays the foundation for such research on noise-induced escape in a heterogeneous system.

\section*{Methods}
\subsection*{\mfyisystem}
In this subsection, we rewrite the system~\eqref{eq:sde-w/mf} in terms of mean field $X$ and displacements $y_i \coloneq x_i - X$.
From equations~\eqref{eq:sde-w/mf} and \eqref{eq:mfdef}, one can write
\begin{align}
  \label{eq:mfyisystem-x}
  \odv{X}{t} =& \frac{1}{N} \sum_{j = 1}^{N} f(X + y_j) + \frac{\alpha}{N} \sum_{j = 1}^{N} \eta_j, \\
  \label{eq:mfyisystem-yi}
  \odv{y_i}{t} =& f(X + y_i) - \frac{1}{N} \sum_{j = 1}^{N} f(X + y_j) - K y_i + \frac{\alpha}{N} \ab[\ab(N - 1) \eta_i - \sum_{j \neq i} \eta_j], \quad i \in \ab\{1, \dots, N - 1\}
\end{align}
where $\eta_i$ are independent white gaussian noise with zero mean and unit variance.

In the drift terms in equations~\eqref{eq:mfyisystem-x} and \eqref{eq:mfyisystem-yi}, we expand $f(X + y)$ as 
\begin{equation}
  f(X + y) = f(X) + f'(X) y + \order{y^2}
\end{equation}
and neglect $\order{y^2}$ to obtain
\begin{gather} \label{eq:drift-mf}
  \frac{1}{N} \sum_{j = 1}^{N} f(X + y_j) \simeq f(X), \\ \label{eq:drift-yi}
  f(X + y_i) - \frac{1}{N} \sum_{j = 1}^{N} f(X + y_j) \simeq f'(X) y_i.
\end{gather}
As for the noise terms, from the reproductive property of gaussian distribution, one can rewrite the terms as
\begin{gather} \label{eq:noise-mf}
  \frac{\alpha}{N} \sum_{j = 1}^{N} \eta_j = \frac{\alpha}{\sqrt{N}} \eta_X, \\ \label{eq:noise-yi}
  \frac{\alpha}{N} \ab[\ab(N - 1) \eta_i - \sum_{j \neq i} \eta_j] = \alpha \sqrt{\frac{N - 1}{N}} \tilde{\eta}_i,
\end{gather}
where $\eta_X$ and $\tilde{\eta}_i$ independently follow $\mathcal{N}(0, 1)$.
The variables $\ab(x_i)$ and $\ab(X, y_i)$ has the following relation,
\begin{equation}
  \begin{pNiceMatrix}
    X \\ y_1 \\ \vdots \\ y_{N - 1}
  \end{pNiceMatrix} 
  = \frac{1}{N} \begin{pNiceMatrix}
    1 & 1 & \cdots & 1 \\
    N - 1 & -1 & \cdots & -1 \\
    \vdots & \ddots & \ddots & \vdots \\
    -1 & \cdots & N - 1 & -1
  \end{pNiceMatrix}
  \begin{pNiceMatrix}
    x_1 \\ x_2 \\ \vdots \\ x_N
  \end{pNiceMatrix}.
\end{equation}
Since the inner product of the first and subsequent rows of the matrix equals to zero, $X$ and $y_i$, and similarly $\eta_X$ and $\tilde{\eta}_i$, are independent.
Putting equations~(\ref{eq:drift-mf}, \ref{eq:drift-yi}, \ref{eq:noise-mf}, \ref{eq:noise-yi}) into equations~\eqref{eq:mfyisystem-x} and \eqref{eq:mfyisystem-yi}, we obtain 
\begin{align}
  %\label{eq:mfyireduced-x}
  \odv{X}{t} =& f(X) + \frac{\alpha}{\sqrt{N}} \eta_X, \\
  %\label{eq:mfyisreduced-yi}
  \odv{y_i}{t} =& -\ab[K - f'(X)] y_i + \alpha \sqrt{\frac{N - 1}{N}} \tilde{\eta}_i, && i \in \ab\{1, \dots, N - 1\},
\end{align}
i.e. equations~\eqref{eq:mfyireduced-x} and \eqref{eq:mfyisreduced-yi}.

\subsection*{\limitingcases}
When there is no interaction, i.e. $K = 0$, $\tau_i$ are independent samples from the distribution for escape times, which has an exponential tail in the weak noise limit~\cite{berglund2013Kramers,creaser2018Sequential}.
Thus $\aet$ is a sample average of $\tau_i$ that are independently sampled from the asymptotic exponential distribution, whose mean is $T_0$ [equation~\eqref{eq:t0}].
The variance of the exponential distribution is ${T_0}^2$.
Furthermore, since the mean escape time $\met$ is the average of sample averages, $\aet$ would follow a normal distribution with mean $T_0$ and variance ${T_0}^2 / N$ due to the central limit theorem, if the sample size $N$ is large enough.
% $\mathcal{N}(T_0, {T_0}^2 / N)$
This expectation was confirmed in figure~\ref{fig:res-all}(d).

% K to infty limit
In the case of infinitely strong coupling, $K \to \infty$, since the system is effectively one-dimensional, the average escape time would follow an exponential distribution whose mean is $T_{\infty}(N)$ [equation~\eqref{eq:tinfty}] in the limit of $K \to \infty$.
As expected, the numerically obtained distribution of average escape times had an exponential tail [figure~\ref{fig:res-all}(f)].

\subsection*{\doublefpe}
The Fokker-Planck equation (FPE) for the uncoupled model~\eqref{eq:sde-uncoupled} is 
\begin{gather} \label{eq:fpe-0}
  \pdv{p_0(t, x)}{t} = -\pdv{}{x} \ab[f(x) p_0(t, x)] + \frac{\alpha^2}{2} \pdv[order=2]{p_0(t, x)}{x},
\end{gather}
where $p_0(t, x)$ is the probability density function (PDF) for $x$ at time $t$.
Using $p_0(t, x)$, the mean field in the uncoupled case, $X_0(t)$, can be calculated as
\begin{gather} \label{eq:x0-fpe}
  X_0(t) = \int_{-\infty}^{\infty} x p_0(t, x) \d{x}.
\end{gather}
By replacing $X$ with $X_0$ in the model SDE~\eqref{eq:sde-w/mf}, the model becomes
\begin{gather}
  \d{x} = \bab{f(x) + K \pab[big]{X_0(t) - x}} \d{t} + \alpha \d{W(t)},
\end{gather}
whose FPE is
\begin{align} \label{eq:fpe-k}
  \pdv{p(t, x)}{t} &= -\pdv{}{x} \bab[Big]{f(x) - K \pab[big]{X_0(t) - x}} p(t, x) + \frac{\alpha^2}{2} \pdv[order=2]{p(t, x)}{x} \\
  &\eqcolon -\pdv{J(t, x)}{x},
\end{align}
where we defined the probability current $J(t, x)$.
$J(t, x)$ represents the probability density that passes through $x$ at time $t$.
We expect the probability density mostly moves towards the upper state around $x = \xi$ because of the strong asymmetry of the potential, i.e. small $r$.
Therefore we may regard $J(t, \xi)$ as the PDF for the escape time $\tau_i$.
That is, $J(t, \xi)$ is the probability that a node escapes at time $t$.
Using $J(t, \xi)$, we can compute the mean and variance of escape times as follows:
\begin{gather}
  \label{eq:met-from-j}
  \met = \mathbb{E}[\tau_i] = \int_0^{\infty} t J(t, \xi) \d{t}, \\
  \label{eq:var-from-j}
  \mathrm{Var}[\tau_i] = \mathbb{E}[\tau_i^2] - \pab{\met}^2 = \int_0^{\infty} t^2 J(t, \xi) \d{t} - \pab{\met}^2
\end{gather}

To obtain our results, we numerically solved FPEs~\eqref{eq:fpe-0} and \eqref{eq:fpe-k} simultaneously to calculate the PDF $p(t, x)$, from which the probability current $J(t, x)$ was computed.
We assumed both $p = 0$ and $\pdif{x} p = 0$ at the boundaries $x = a$ and $b$.
We used the central difference to approximate the space derivative with step size $\Delta x$, obtaining a system of ordinary differential equations on discretized space.
Integrals such as the ones in equations~\eqref{eq:x0-fpe}, \eqref{eq:met-from-j}, and \eqref{eq:var-from-j} were approximated by summations within finite sections.

\subsection*{\nummethod}
\begin{table}[tb]
  \centering
  \caption{%
    A list of parameters that appear in this paper.
    Specific values of relevant parameters are indicated in each figure.
  }
  \label{tab:paramlist}
  \small
  \begin{tabularx}{\linewidth}{rX}
    \toprule
    \textbf{parameter} & \textbf{description} \\ \midrule
    $r$ & The location of the potential barrier. \\
    $\alpha$ & The noise intensity. \\
    $N$ & The number of nodes. \\
    $K$ & The coupling strength. \\
    $\xi$ & The threshold to determine the escape time. \\
    $\rmsub{n}{trial}$ & The number of trials, i.e. sample paths. \\
    dt & The step size in time used in the Euler-Maruyama method. \\
    $\Delta t$ & The time interval to sample results of numerical integration. \\
    seed & The seed for a random number generator. \\
    $a$ & The lower end of the state space to solve Fokker-Planck equations. \\
    $b$ & The upper end of the state space to solve Fokker-Planck equations. \\
    $\Delta x$ & The step size in space to discretize space derivatives in Fokker-Planck equations. \\
    $\rmsub{X}{end}$ & Numerical integration of Fokker-Planck equations was terminated when the mean field $X$ exceeded $\rmsub{X}{end}$. \\
    \bottomrule
  \end{tabularx}
\end{table}

Table~\ref{tab:paramlist} is the list of parameters that appear in this paper.
We noted specific parameter values used to obtain the corresponding results in each figure so as to eliminate room for a mistake in transcribing values into the manuscript.
For numerical integration of SDEs, we implemented the Euler-Maruyama method with fixed time step dt.
The results were sampled with the time interval of $\Delta t$ due to memory limitation.
To obtain first escape times numerically, we simulated the model SDE [equation~\eqref{eq:sde-w/mf}] from the initial condition of $x_i = 0$ for all $i$.
We say node $i$ has escaped when $x_i \geq \xi$ holds for a given threshold $\xi$.
Each simulation run was terminated when all nodes had escaped, and the first time step when $x_i \geq \xi$ was satisfied was recorded as $\tau_i$ for each node.
For each parameter value, this process was repeated $\rmsub{n}{trial}$ times, using different seeds for the random number generator.
For numerical integration of ordinary differential equations, we used \texttt{scipy.integrate.solve\_ivp()} method with Python.
The parameters regarding the error tolerance, \texttt{a\_tol} and \texttt{r\_tol}, were both set to $10^{-8}$.
When solving the Fokker-Planck equations in our approximate theory, the results were sampled with the time interval of $\Delta t$ due to memory limitation.
This was done by assigning \texttt{t\_eval} argument so that the solver interpolated values at $t = j \, \Delta t$ ($j \in \mathbb{N}$).

\section*{Data Availability}
The data and scripts used in this research are available in the GitHub repository, \url{https://github.com/ishiihidemasa/24-coupling-facilitate-impede-escape}.

\bibliography{24-coupling-facilitate-impede}

\section*{Acknowledgements}
We thank Istvan Z. Kiss, Hajime Koike, Naoki Masuda, and Yuzuru Sato for valuable comments.
This study was supported by the World-leading Innovative Graduate Study Program in Proactive Environmental Studies (WINGS-PES), the University of Tokyo, to H.I., and JSPS KAKENHI (No. JP21K12056) to H.K.

\section*{Author contributions statement}
H.I. and H.K. conceived this project.
H.I. conducted numerical simulations, analyzed the results, and wrote the manuscript in consultation with H.K.

\section*{Additional information}
\paragraph*{Competing interests} The authors declare no competing interests.

\end{document}